\documentclass[runningheads]{llncs}
\usepackage[T1]{fontenc}
\usepackage{graphicx}
\usepackage{caption,subcaption}
\usepackage{multirow}
\usepackage[table]{xcolor}
\usepackage{url}
\usepackage{setspace}
\usepackage{color}
\begin{document}
\title{Benchmarking Early Agitation Prediction in Community-Dwelling People with Dementia Using Multimodal Sensors and Machine Learning}
\titlerunning{Agitation Prediction in Community-Dwelling People with Dementia}

\author{
Ali Abedi\inst{1,2}\orcidID{0000-0002-7393-1362} \and
Charlene H. Chu\inst{1,2}\orcidID{0000-0002-0333-7210} \and
Shehroz S. Khan\inst{1,3}\orcidID{0000-0002-1195-4999}
}

\authorrunning{Abedi et al.}

\institute{KITE Research Institute, Toronto Rehabilitation Institute, University Health Network, Toronto, Canada 
\and
Lawrence Bloomberg Faculty of Nursing, University of Toronto, Toronto, Canada
\and
College of Engineering and Technology, American University of the Middle East, Egaila, Kuwait\\
\email{\{ali.abedi,shehroz.khan\}@uhn.ca, charlene.chu@utoronto.ca}
}

\maketitle              
\begin{abstract}

Agitation is one of the most common responsive behaviors in people living with dementia, particularly among those residing in community settings without continuous clinical supervision. Timely prediction of agitation can enable early intervention, reduce caregiver burden, and improve the quality of life for both patients and caregivers. This study aimed to develop and benchmark machine learning approaches for the early prediction of agitation in community-dwelling older adults with dementia using multimodal sensor data. A new set of agitation-related contextual features derived from activity data was introduced and employed for agitation prediction. A wide range of machine learning and deep learning models was evaluated across multiple problem formulations, including binary classification for single-timestamp tabular sensor data and multi-timestamp sequential sensor data, as well as anomaly detection for single-timestamp tabular sensor data. The study utilized the Technology Integrated Health Management (TIHM) dataset, the largest publicly available dataset for remote monitoring of people living with dementia, comprising 2,803 days of in-home activity, physiology, and sleep data. The most effective setting involved binary classification of sensor data using the current 6-hour timestamp to predict agitation at the subsequent timestamp. Incorporating additional information, such as time of day and agitation history, further improved model performance, with the highest AUC-ROC of 0.9720 and AUC-PR of 0.4320 achieved by the light gradient boosting machine. This work presents the first comprehensive benchmarking of state-of-the-art techniques for agitation prediction in community-based dementia care using privacy-preserving sensor data. The proposed approach demonstrates that accurate, explainable, and efficient agitation prediction is feasible, paving the way for proactive dementia care solutions that support aging in place.

\end{abstract}
\section{Introduction}
\label{sec:introduction}
Dementia is a progressive neurodegenerative condition that affects millions globally, posing substantial challenges for individuals, families, and healthcare systems~\cite{livingston2024dementia}. As populations age, the number of people living with dementia (PLwD) continues to rise rapidly~\cite{livingston2024dementia}. According to the World Health Organization, over 55 million people currently live with dementia worldwide, a figure projected to reach 78 million by 2030 and 139 million by 2050~\cite{who2023dementia}. This growing prevalence underscores the urgent need for scalable, proactive solutions to manage dementia-related clinical and behavioral challenges, particularly as more PLwD continue to reside in their homes or community settings rather than institutional care~\cite{cihi2018dementia}.

Responsive behaviors in PLwD, such as aggression, wandering, agitation, and resistance to care, often result from unmet needs, environmental stressors, or health issues. Agitation is especially common, presenting as verbal or physical aggression, restlessness, or emotional distress, and affecting up to 70\% of PLwD~\cite{carrarini2021agitation}. It can increase caregiver burden, speed cognitive decline, and lead to early institutionalization and poorer quality of life~\cite{cloutier2019institutionalization,schein2022impact}. While many studies focus on managing agitation in long-term care~\cite{wong2024barriers,khan2024novel,mishra2023privacy,deters2024sensor}, over 60\% of PLwD live in private homes~\cite{cihi2018dementia}. Still, agitation in these settings is often under-monitored due to a lack of real-time support and early warning tools~\cite{palermo2023tihm}.

Timely prediction of agitation can offer significant benefits in dementia care. Forecasting episodes allows caregivers to implement personalized calming strategies, adjust environments, or intervene early to prevent escalation~\cite{hekmatiathar2021data,anderson2021dementia}. Early prediction may reduce emergency visits, delay institutionalization, and improve the well-being of both PLwD and caregivers~\cite{gitlin2012nonpharmacologic}. However, accurately identifying agitation risk remains challenging due to symptom variability, limited monitoring, and a lack of predictive tools in community settings~\cite{palermo2023tihm}.

Monitoring mobility, physiological signals such as blood pressure and heart rate, and sleep patterns can greatly improve agitation management, detection, and prediction, given their established associations with physical activity, autonomic changes, and sleep disruptions in PLwD~\cite{cheung2022wrist,liu2023heart,de2021association,brown2015sleep,iaboni2022wearable}. The emergence of Internet of Things (IoT), wearable sensors, and Artificial Intelligence (AI) offers a promising direction~\cite{abedi2024artificial,abedi2022maison,iaboni2022wearable,sheikhtaheri2022applications}. IoT devices such as wearables, smartphones, and ambient sensors can passively collect continuous physiological and behavioral data. These multimodal streams capture daily routines, sleep, and mobility patterns that may precede agitation. Machine learning techniques can analyze this data to uncover hidden patterns and predict agitation risk~\cite{deters2024sensor,khan2018detecting,hekmatiathar2021data}. The integration of IoT sensing and AI modeling enables scalable, personalized, and non-invasive agitation monitoring platforms.

While machine learning offers promise for managing agitation in dementia, most research has focused on detecting agitation as it occurs rather than predicting it in advance~\cite{khan2018detecting,palermo2023tihm,mishra2023privacy,iaboni2022wearable}. This paper distinguishes between detection, reactive and limited, and prediction, which enables timely intervention. Most prior studies have been conducted in institutional settings~\cite{wong2024barriers,khan2024novel,mishra2023privacy,deters2024sensor}, limiting relevance to PLwD living in the community. As a result, a key gap remains in developing privacy-preserving predictive tools tailored for community-dwelling PLwD.

To address these gaps, this study explores the early prediction of agitation in community-dwelling PLwD using multimodal, privacy-preserving sensor data. The proposed method leverages data from mobility and physiological IoT devices to train machine learning and deep learning models capable of forecasting agitation episodes before they occur. This study makes the following contributions:
\begin{itemize}
    \item A new set of contextual features related to agitation was derived from activity data and utilized for agitation prediction. Additional contextual information, including time of day and agitation history, was incorporated to enhance prediction performance.
    \item A diverse set of machine learning and deep learning models was evaluated across various problem formulations, including binary classification and anomaly detection for single-timestamp tabular sensor data, and binary classification for multi-timestamp sequential sensor data.
    \item Extensive experiments were conducted across different settings on a publicly available dataset, with explainability incorporated through analysis of feature importance and contribution.
\end{itemize}

By predicting agitation before it occurs and using non-intrusive in-home sensors, this study offers a scalable, privacy-conscious framework for supporting PLwD. The results demonstrate the potential of AI-driven remote monitoring to enable proactive dementia care~\cite{waldman2019healthcare}.


\section{Related Work}
\label{sec:related_work}
This section provides an overview of existing research on agitation detection and prediction in PLwD using multimodal sensors and machine learning and deep learning algorithms.

Sensor-based agitation detection and prediction in PLwD was first reviewed by Khan et al.\cite{khan2018detecting}, who demonstrated feasibility but noted a reliance on single-modality actigraphy data. Deters et al.\cite{deters2024sensor} later provided a broader review of institution-based studies, categorizing them under three platforms: InsideDEM~\cite{teipel2017multidimensional}, ORCATECH~\cite{kaye2018methodology}, and DAAD~\cite{khan2017daad}. While InsideDEM and ORCATECH supported association analyses between agitation and variables such as mobility, sleep, and heart rate, DAAD enabled predictive model development. These platforms employed various sensors, including video, wearables, and ambient devices, to capture agitation’s multifactorial nature. Some of the existing approaches are detailed below.

Khan et al. developed the DAAD platform~\cite{khan2017daad} and conducted one of the earliest studies using multimodal sensors for agitation detection~\cite{khan2019agitation}. Data from two PLwD over 28 days were collected via Empatica E4, showing that multimodal inputs outperformed single sensors using a Random Forest (RF) classifier. Spasojevic et al.\cite{spasojevic2021pilot} extended this work with 600 days of data from 17 PLwD, confirming the value of multimodal sensing. To address class imbalance towards non-agitation episodes, Meng et al.\cite{meng2024undersampling} introduced a hybrid undersampling and class re-decision method using sequential patterns to improve performance. Iaboni et al.~\cite{iaboni2022wearable} applied feature ranking and developed personalized models, revealing individual differences in feature importance and benefits of individualized detection approaches.

Apart from the studies based on the DAAD platform~\cite{khan2017daad}, several works have investigated alternative approaches to agitation detection using vocal features and in-home sensors. Salekin et al.\cite{salekin2017dave} developed a system for identifying different types of verbal agitation, through a combination of acoustic signal processing and text mining with language models, achieving high detection accuracy. Rezvani et al.\cite{rezvani2021semi} collected multimodal sensor data from the homes of 96 PLwD using smart plugs, motion sensors, and door sensors, and developed machine learning models to estimate agitation likelihood. To address limited labeled data, they employed a two-stage semi-supervised strategy: a self-supervised transformation learning model generated pseudo-labels from unlabeled data, followed by a Bayesian ensemble trained on labeled samples. This method outperformed Long Short-Term Memory (LSTM) and Support Vector Machine (SVM) baselines in terms of Area Under the Receiver Operating Characteristic Curve (AUC-ROC).

Palermo et al. \cite{palermo2023tihm} utilized the Technology Integrated Health Management (TIHM) dataset to develop machine learning models for agitation detection in community-dwelling PLwD, based on statistical features extracted from multimodal sensor data. Among the evaluated models, Logistic Regression (LR) and Gaussian Process classifiers achieved the highest sensitivity and specificity.

Badawi et al.~\cite{badawi2024leveraging} proposed a semi-supervised learning framework for detecting agitation in PLwD using wearable physiological sensor data. Data were collected from 14 participants across three hospitals using the Empatica E4 wristband, which recorded heart rate, acceleration, electrodermal activity, and skin temperature. With only five participants having fully labeled agitation episodes, the study employed a hybrid approach combining variational autoencoders (VAE) for latent feature extraction and a self-training strategy to iteratively generate pseudo-labels for unlabeled data. The best performance was achieved using XGBoost within the self-training framework enhanced by VAE-derived features.

A limited number of studies have focused on predicting agitation before it occurs. Homdee~\cite{homdee2020prediction} collected environmental data, such as light, temperature, humidity, air pressure, and acoustic noise, from three PLwD over two months. Statistical features were used to train gradient boosting machines and LSTM models, which showed moderate performance, indicating feasibility and room for improvement. Similarly, HekmatiAthar et al.~\cite{hekmatiathar2021data} monitored one community-dwelling PLwD over 64 days using sensors that captured ambient noise, illuminance, temperature, pressure, and humidity. An LSTM model was trained to predict agitation, with downsampling applied to address class imbalances towards non-agitation moments.

Based on the reviewed literature, the majority of existing studies have focused primarily on agitation detection, identifying agitation as it occurs \cite{khan2018detecting,deters2024sensor,khan2017daad,khan2019agitation,spasojevic2021pilot,meng2024undersampling,iaboni2022wearable,salekin2017dave,rezvani2021semi,palermo2023tihm,badawi2024leveraging}, rather than on its prediction prior to onset. Moreover, much of this research has been conducted in controlled clinical or institutional settings, limiting the generalizability of findings to real-world, home-based environments. Among the relatively few studies that have addressed agitation prediction \cite{homdee2020prediction,hekmatiathar2021data}, most were constrained by small sample sizes or limited sensor modalities, thereby restricting their applicability to scalable deployment. In contrast, the present study explores predicting agitation before it occurs. Leveraging a large, real-world, home-based, and multimodal dataset, this work advances the field by addressing key gaps in prediction timing and scalability, with the goal of supporting proactive and personalized dementia care in home settings.

\section{Technology Integrated Health Management (TIHM) Dataset}
\label{sec:dataset}
\subsubsection{Participants}
\label{sec:participants}
The TIHM dataset \cite{palermo2023tihm} was collected from 56 PLwD residing in home settings. It comprises a total of 2,803 days of data. The duration of data collection per participant ranged from 3 to 90 days, with a mean of 48.91 days and a standard deviation of 23.09 days. Among the participants, 50\% were female; 30\% were aged 70–80, 47\% were 80–90, and 23\% were 90–100 years old. Additionally, 89\% identified as white, and 25\% were living alone.

\subsubsection{Data}
\label{sec:data}
TIHM captures a rich set of digital biomarkers through a combination of ambient and physiological sensing modalities.
\begin{itemize}
    \item \textbf{Activity--}Continuous passive infrared (PIR) motion sensors were installed across eight locations within participants’ homes, including back-door, bathroom, bedroom, fridge-door, front-door, hallway, kitchen, and lounge, capturing ambient motion on an event-driven basis whenever movement occurred
    \item \textbf{Physiology--}In parallel, a suite of physiological sensors was used by participants to measure eight health parameters multiple times per day, including body-temperature, body-weight, diastolic and systolic blood pressure, heart-rate, muscle-mass, total-body-water, and skin-temperature.
    \item \textbf{Sleep--}The dataset also includes detailed sleep data, capturing heart rate and respiratory rate during awake, light, deep, and rapid eye movement sleep episodes.
\end{itemize}
Out of the 2,803 days of data available in TIHM, the proportions of missing data calculated on a daily basis were 2.89\% for activity, 22.90\% for physiology, and 70.21\% for sleep. Notably, 69.64\% of participants had no recorded sleep data.

\subsubsection{Label}
\label{sec:label}
TIHM provides clinician-verified labels at 6-hour time resolutions for six key clinical events: agitation episodes, abnormal blood pressure (high or low), abnormal body temperature (high or low), dehydration (low body water), abnormal heart rate (high or low), and significant weight changes. The availability of labels aligned with the multimodal sensor data makes the dataset well-suited for supervised machine learning and deep learning model development.

\subsubsection{Agitation}
Of the 56 participants in TIHM, 27 experienced at least one agitation episode, while 29 did not exhibit any agitation during the monitoring period. In total, 135 agitation episodes were recorded, with the number of episodes per participant ranging from 1 to 33, measured in 6-hour intervals. Across all participants, the mean and standard deviation of agitation episodes were 2.41 and 5.26, respectively. Among those who experienced at least one agitation episode, the mean and standard deviation were 5.00 and 6.71, respectively. Although some participants exhibited multiple agitation episodes within a single day, most episodes were temporally scattered. On a per-participant basis, the mean and standard deviation of the interval between consecutive agitation episodes were 2.43 days and 4.15 days, respectively.

While no agitation episodes were reported before 6:00 a.m., 8.15\% occurred between 6:00 a.m. and 12:00 p.m., 58.52\% between 12:00 p.m. and 6:00 p.m., and 33.33\% between 6:00 p.m. and midnight. The increased occurrence of agitation in the afternoon and evening aligns with what is commonly referred to as the sundowning phenomenon in PLwD \cite{canevelli2016sundowning}, during which individuals may experience heightened confusion, restlessness, or distress later in the day.

\section{Method}
\label{sec:method}
This section presents the proposed pipeline for agitation prediction in PLwD, outlining the preprocessing, feature extraction, problem formulation, and predictive modeling components.

\subsection{Preprocessing}
\label{sec:preprocessing}
Preprocessing involved missing value imputation, standardization, and normalization. As detailed in section~\ref{sec:data}, the sleep data modality exhibited a high degree of missingness and was therefore excluded from this study, as it was deemed unsuitable for imputation and subsequent predictive modeling \cite{madley2019proportion,junaid2025much}. Consequently, activity and physiology data were used for agitation prediction. Missing values were filled using the mean from the training set, applied to both training and test sets. Data were then standardized and normalized, with all transformations fitted on the training data and applied consistently.

\subsection{Feature Extraction}
\label{sec:feature_extraction}
The raw activity data consist of motion detection events recorded with second-level accuracy across eight home locations. The first set of activity features includes 32 statistical variables, derived as described in~\cite{palermo2023tihm}. For each location, hourly activity counts are first computed. Then, for each timestamp (e.g., a 6-hour period), the sum, maximum, mean, and standard deviation of these hourly counts are calculated. Each feature is labeled using the format 'location-count-statistic' (e.g., \textit{hallway-count-sum}).

The second set of activity features comprises eight contextual features, each calculated over a given timestamp (e.g., a 6-hour period) as follows: \textit{total-events}, the total number of detected activities across all monitored locations, reflecting the overall level of movement or environmental interaction; \textit{unique-locations}, the number of distinct locations where activity was detected, indicating the range of spatial engagement within the home; \textit{active-location-ratio}, the proportion of monitored locations that recorded at least one activity event, providing a normalized measure of spatial activity diversity; \textit{private-to-public-ratio}, the ratio of activity events detected in private areas (e.g., bedroom, bathroom) to those in public or shared areas (e.g., kitchen, hallway, lounge), offering insight into the participant’s preference for private versus public spaces; \textit{location-entropy}, a measure of the randomness or unpredictability in the participant’s spatial behavior. Higher values indicate more evenly distributed activity across locations, while lower values suggest concentration in a limited number of areas; \textit{location-dominance-ratio}, the proportion of activity events that occurred in the most active location, capturing the degree to which behavior was dominated by a single area; \textit{back-and-forth-count}, the number of instances where a participant returned to a previously visited location after briefly moving to another, potentially indicating restlessness or repetitive movement patterns; \textit{num-transitions}, the total number of transitions between different locations, serving as a measure of mobility and spatial movement complexity.

For each timestamp (e.g., a 6-hour window) in TIHM, there are typically a few, sometimes only one, recorded values for each of the eight physiological measurements. The corresponding eight physiology features are computed as the mean of all available measurements within each timestamp. Therefore, a total of 48 features are extracted from the activity and physiology data at each timestamp, comprising 32 statistical activity features, 8 contextual activity features, and 8 physiology features.

\subsection{Agitation Prediction Problem Formulation}
\label{sec:formulation}
This section presents the proposed formulations for agitation prediction, including binary classification using tabular and sequential sensor data, as well as anomaly detection using tabular sensor data.

\subsubsection{Binary Classification}
\label{sec:binary_classification}
The goal is to predict agitation at the next timestamp ($t + 1$), using sensor data from the previous $n$ timestamps ($t - (n - 1), \dots, t$), based on the cyclic nature of dementia symptoms~\cite{rak2024prevalence,bachman2006sundowning,cohen2007temporal}. For $n = 1$, each sample is a single feature vector at time $t$, forming a tabular dataset suitable for standard machine learning models. For $n > 1$, each sample becomes a sequence of $n$ feature vectors, requiring sequential models designed to handle temporal dependencies.


\subsubsection{Anomaly Prediction}
\label{sec:anomaly_prediction}
While agitation detection has been widely explored as an anomaly detection task~\cite{mishra2023privacy,9684388}, this study examines agitation prediction as an anomaly prediction problem. Here, a sequence of $n$ past timestamps is labeled \textit{normal} if agitation does not occur at $t + 1$, and \textit{anomalous} otherwise. Models are trained exclusively on normal samples to learn the underlying distribution of non-agitated behavior. During inference, deviations from this learned distribution indicate a higher likelihood of agitation in the next timestamp.

\subsection{Predictive Modeling}
\label{sec:predictive_modeling}
A range of machine learning and deep learning models were employed for agitation prediction, based on the problem formulations described in subsection \ref{sec:formulation}. The models designed for single-timestamp tabular data included a Transformer-based model for tabular data, named Tabular Prior-data Fitted Network (TabPFN) \cite{hollmann2025accurate} as well as gradient boosting decision tree algorithms, including Gradient Boosting (GB) classifier \cite{friedman2001greedy} and Light Gradient Boosting Machine (LightGBM) \cite{ke2017lightgbm}. In addition, traditional machine learning classifiers were evaluated \cite{pranckevivcius2017comparison}, including LR and Naive Bayes (NB). The models employed for multi-timestamp sequential data included the Transformer architecture \cite{vaswani2017attention} and ROCKET (RandOm Convolutional KErnel Transform) \cite{dempster2020rocket}. Finally, the models used for anomaly prediction on single-timestamp tabular data included One-Class SVM, Isolation Forest (IF), and Local Outlier Factor (LOF) \cite{ahmad2025one}.

Due to the highly imbalanced distribution of data samples in the dataset, with agitation, the class of interest, being the minority, class imbalance was addressed using weighted loss functions \cite{cui2019class} and the Synthetic Minority Over-sampling Technique (SMOTE) \cite{chawla2002smote}. The weights in the loss functions were set proportional to the class distribution, ensuring greater penalization for misclassifying minority class instances.

\section{Experiments}
\label{sec:experiments}
This section presents the experimental settings and results of the proposed method across different settings for agitation prediction in PLwD.

\subsubsection{Agitation Prediction as Binary Classification--Tabular Sensor Data}
\label{sec:agitation_prediction_tabular}
Table 1 presents results for agitation prediction as a binary classification problem, identifying whether agitation occurs at $t + 1$ using data from $t$ ($n = 1$, per subsection~\ref{sec:formulation}), based on 6-hour samples from TIHM. Models were trained on statistical and contextual activity and physiology features, with and without weighted loss functions, and evaluated using both 5-fold CV and leave-one-participant-out (LOPO) CV settings.

Table 1 shows that both AUC-ROC and AUC-PR values significantly surpass the baseline performance of a random binary classifier. Weighted models generally outperform their unweighted counterparts. As expected, LOPO results are lower than those from 5-fold CV but remain comparable, indicating generalization to unseen participants. The Transformer (TabPFN) and LR models achieve the highest AUC-ROC across different settings, while NB attains the highest AUC-PR. Although NB and LR detect the most agitation episodes, as reflected in their high sensitivity, they also produce more false positives, resulting in lower specificity. Due to the relatively better performance of models trained with weighted loss functions, all subsequent experiments employed this setting.

\subsubsection{Agitation Prediction as Binary Classification--Sequential Sensor Data}
\label{sec:agitation_prediction_sequential}
Figure 1 presents binary classification results using data from the $n$ most recent timestamps as sequential data samples from TIHM. Models include a Transformer encoder and a combination of ROCKET~\cite{dempster2020rocket} and GB~\cite{friedman2001greedy} with a weighted loss function, trained on statistical and contextual activity and physiology features under 5-fold CV. The Transformer consistently outperforms ROCKET, achieving the highest AUC-ROC and AUC-PR at $n = 2$. When more than two past timestamps are used, the Transformer's agitation prediction performance declines.

\begin{center}
    \includegraphics[width=\textwidth]{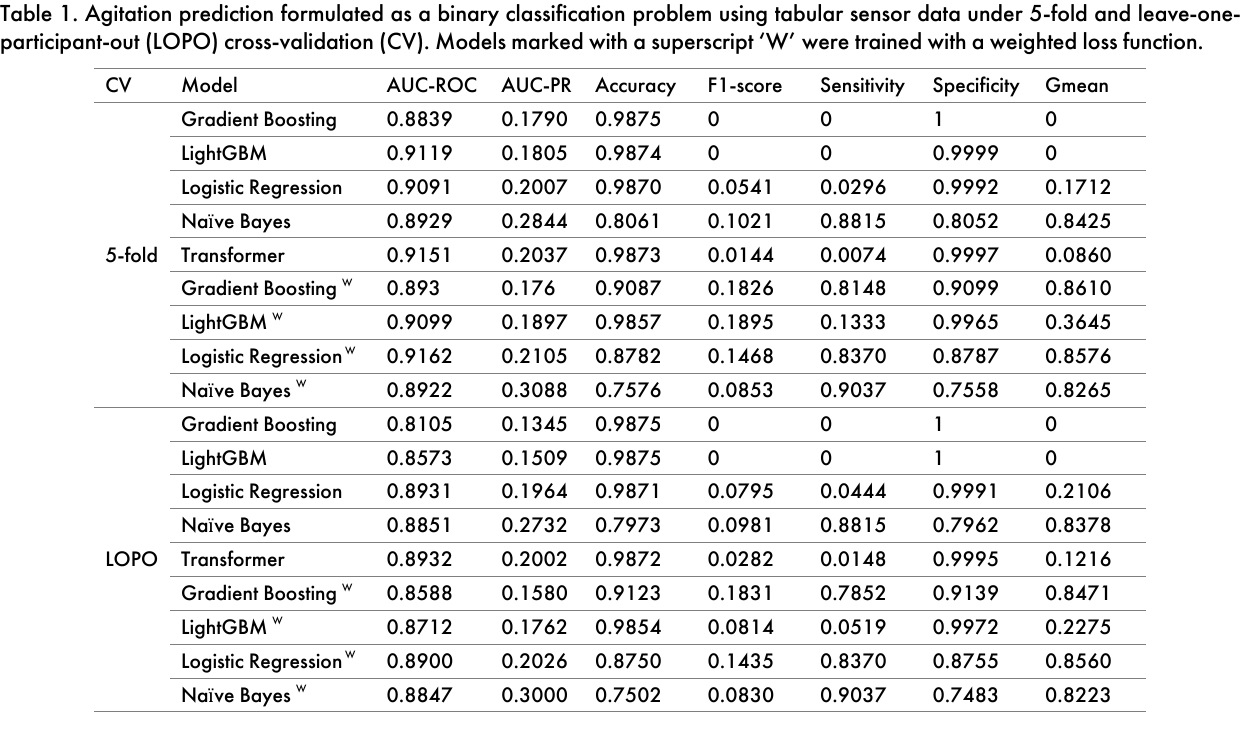}
\end{center}

\begin{center}
    \includegraphics[width=\textwidth]{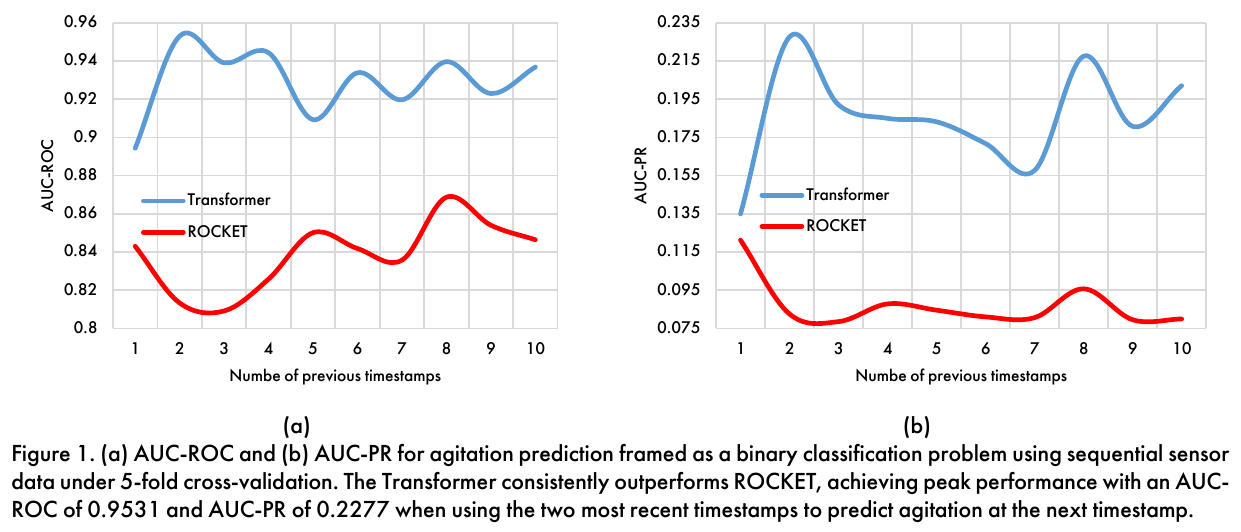}
\end{center}

\begin{center}
    \includegraphics[width=\textwidth]{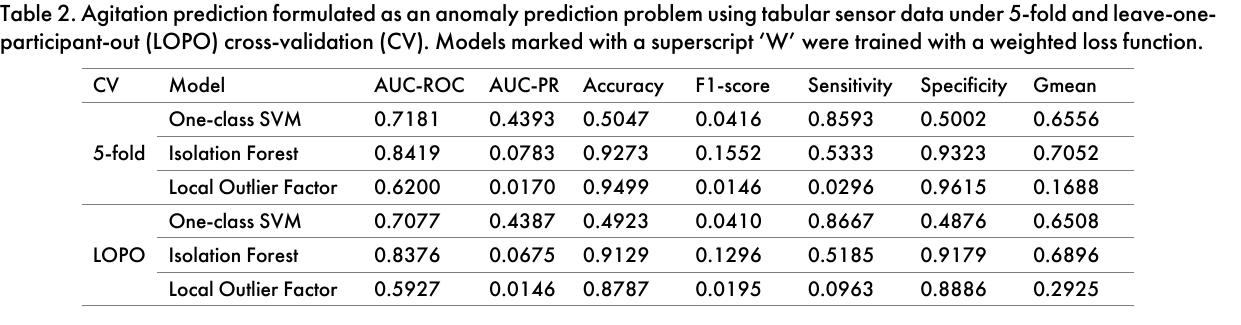}
\end{center}

\subsubsection{Agitation Prediction as Anomaly Prediction}
\label{sec:agitation_prediction_anomaly}
Table 2 presents results for agitation prediction framed as an anomaly prediction problem, using data from the current timestamp ($n = 1$) with 6-hour samples from TIHM, evaluated under 5-fold and LOPO CV. Models were trained only on normal data (non-agitation at the next timestamp). Compared to binary classification (Table 1), anomaly prediction yields lower AUC-ROC overall, though One-class SVM achieves higher AUC-PR than the classification models. These results suggest that while anomaly prediction may struggle with overall discrimination, it can be more effective in identifying rare agitation cases.

\subsection{Temporal Resolution}
\label{sec:temporal_resolution}
Figure 2 shows agitation prediction performance, formulated as binary classification of tabular data across 6-hour, 12-hour, and 24-hour temporal resolutions. The number of positive cases varies slightly by resolution, with samples containing multiple agitation episodes labeled the same as those with only one. Longer intervals result in fewer total samples, reducing available negative cases for training and testing. The 6-hour setting yields the highest AUC-PR and AUC-ROC, followed by 12-hour and then 24-hour, indicating improved prediction accuracy with shorter temporal windows. These findings underscore the clinical value of higher-frequency monitoring for timely and reliable agitation prediction.

\begin{center}
    \includegraphics[width=\textwidth]{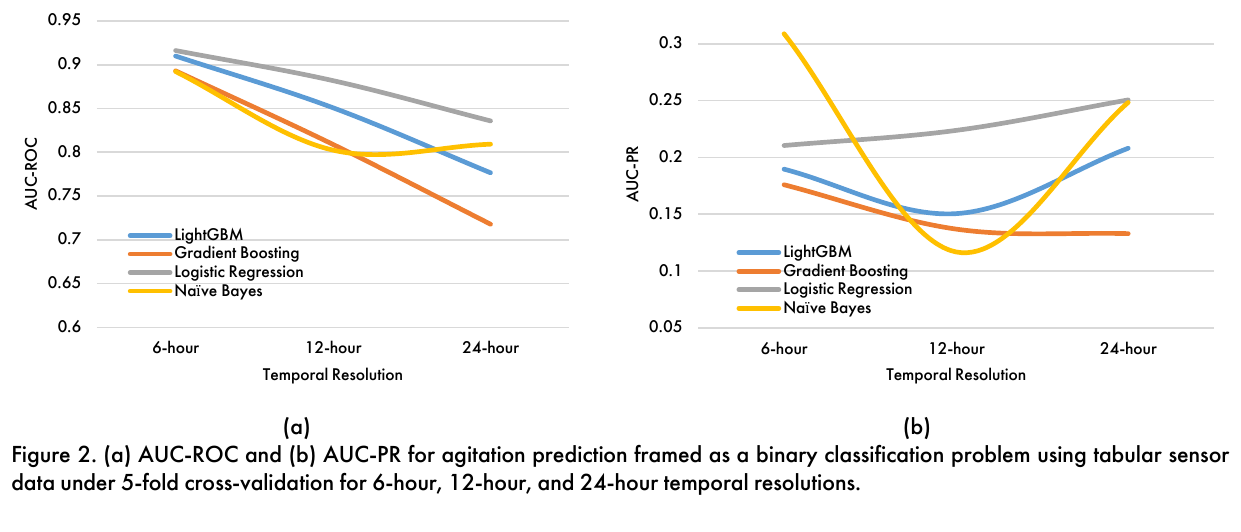}
\end{center}


\begin{center}
    \includegraphics[width=\textwidth]{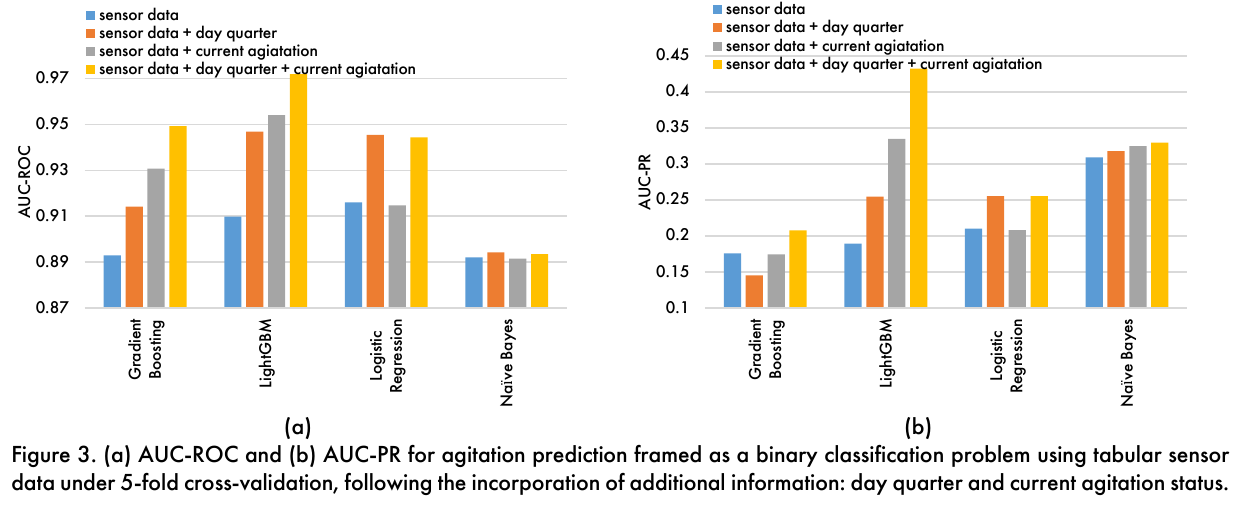}
\end{center}

\subsection{Incorporating Additional Information}
\label{sec:additional_information}
As noted in section~\ref{sec:dataset}, agitation episodes in TIHM were unevenly distributed across the day, with a higher concentration in the afternoon and evening. This pattern reflects the well-known sundowning phenomenon in PLwD~\cite{canevelli2016sundowning}, motivating the inclusion of day-quarter information (encoded as an integer) as an additional input to models alongside activity and physiology features. Temporal analysis also showed that 86.67\% of agitation episodes occurred without agitation in the previous timestamp, while only 13.33\% were preceded by an agitation. This motivated the inclusion of a binary feature indicating current agitation to support the prediction of future agitation. Importantly, incorporating day-quarter and current-timestamp agitation as input features does not introduce information leakage, as both are derived solely from the current or past timestamps and exclude any future data.

Figure 3 presents the results of the proposed method under 5-fold CV with incremental inclusion of additional contextual information. Incorporating each piece of information individually, and especially both together, leads to substantial improvements in AUC-ROC and AUC-PR across all settings, reaching up to 0.9720 and 0.4320, respectively, for LightGBM. These results highlight the value of integrating contextual cues to enhance agitation prediction performance.

\subsection{Feature Importance and Contribution}
\label{sec:feature_importance}
SHapley Additive exPlanations (SHAP) \cite{lundberg2017unified} was used to evaluate the contribution of statistical, contextual activity, and physiology features to agitation prediction. LightGBM, selected for its strong performance in earlier experiments, served as the base model under 5-fold CV. Figure 4 shows the SHAP summary plot of the top 24 features, highlighting both their importance and the effect of low vs. high values on predictions. Features are ranked by impact, with the top feature being a statistical activity measure—hallway-count-std (see section~\ref{sec:feature_extraction}). The predominance of high values (in red) on the positive SHAP axis suggests that greater hallway movement variability increases the likelihood of agitation in the next timestamp. Similar interpretations apply to other features based on SHAP value distribution and color. Notably, several proposed contextual features also rank among the most influential.

\begin{center}
    \includegraphics[width=.9\textwidth]{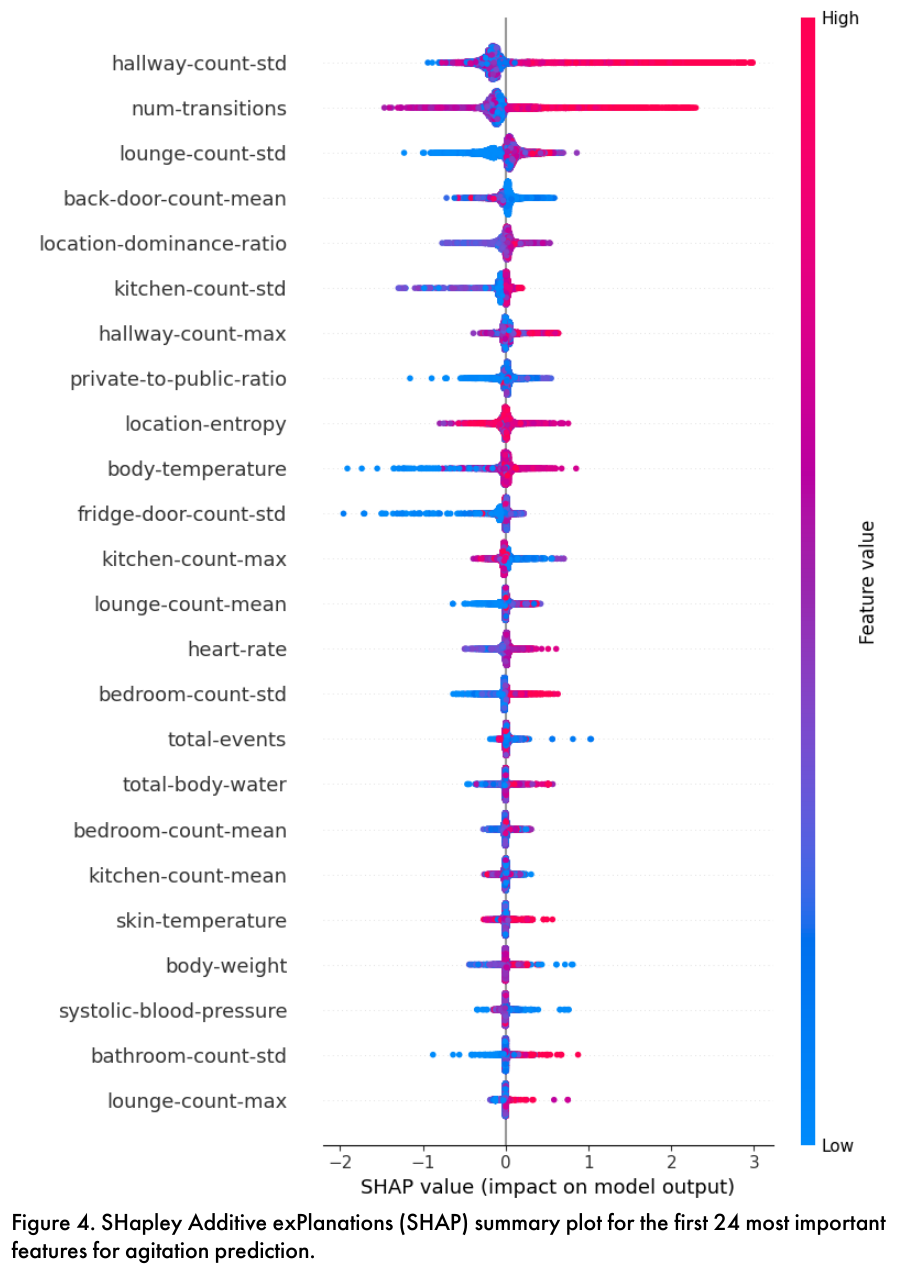}
\end{center}

\section{Conclusion}
\label{sec:conclusion}
This study investigated early prediction of agitation episodes in community-dwelling PLwD using multimodal sensor data and machine learning. Leveraging the TIHM dataset, which contains continuous in-home activity and physiological data, multiple problem formulations were explored. Novel contextual features derived from activity data were introduced for prediction. The binary classification model using the current 6-hour timestamp consistently achieved the best performance, further enhanced by incorporating time-of-day and current agitation status information. A key limitation of this study is the limited number of agitation episodes for model training, which may impact generalizability and lead to lower true positive rates. Although the TIHM dataset provides rich multimodal data, it represents a specific cohort of community-dwelling older adults with dementia, potentially limiting applicability to other settings or populations. The use of fixed 6-, 12-, and 24-hour timestamps may also miss finer temporal patterns. Future work should explore larger, more diverse datasets and adaptive temporal models to enhance prediction robustness and relevance. In conclusion, this study demonstrates the feasibility of accurate and explainable agitation prediction in community-dwelling PLwD using privacy-preserving, non-invasive multimodal sensor data. The results hold significant promise for proactive dementia care by supporting timely interventions and alleviating caregiver burden.

\subsubsection{Acknowledgment} This work was supported by Longitude Prize on Dementia, UK, and the Alzheimer Society of Canada.

\begingroup
\setstretch{1.} 
\bibliographystyle{IEEEtran}
\bibliography{references}
\endgroup

\end{document}